# Open Access in Europe: A National and Regional Comparison

Abdelghani Maddi[1], Esther Lardreau[2], David Sapinho[3]

[1] *abdelghani.maddi@hceres.fr*
Observatoire des Sciences et Techniques, Hcéres, 2 Rue Albert Einstein, Paris, 75013 France.
Sorbonne Paris Nord, CEPN, UMR-CNRS 723, Villetaneuse, 93420 France.

[2] *esther.lardreau@hceres.fr*
Observatoire des Sciences et Techniques, Hcéres, 2 Rue Albert Einstein, Paris, 75013 France

[3] *david.sapinho@hceres.fr*
Observatoire des Sciences et Techniques, Hcéres, 2 Rue Albert Einstein, Paris, 75013 France

## Abstract

Open access to scientific publications has progressively become a key issue for European policy makers, resulting in concrete measures by the different country members to promote its development.

The aim of paper is, after providing a quick overview of OA policies in Europe, to carry out a comparative study of OA practices within European countries, using data from the Web of Science (WoS) database.

This analysis is based on two indicators: the OA share that illustrates the evolution over time, and the normalized OA indicator (NOAI) that allows spatial comparisons, taking into account disciplinary structures of countries.

Results show a general trend towards the development of OA over time as expected, but with large disparities between countries, depending on how early they begin taking measures in favor of OA.

While it is possible to stress the importance of policy and its influence on open access at country level, this does not appear to be the case at the regional level. There is not much variability between regions in terms of open access indicators.

---

[1] Corresponding author : Abdelghani Maddi, Observatoire des Sciences et Techniques, Hcéres, 2 Rue Albert Einstein, Paris, 75013, France, T. 33 (0)1 55 55 61 48.

# Introduction

Over the last decades, the scientific community has been questioning its knowledge diffusion model and has been exploring alternatives to bypass the access restrictions resulting from increasing subscription fees (Houghton 2001; Tananbaum 2003; Chi Chang 2006; Bruns *et al.* 2020). In this context, Open Access (OA) models of scientific publications appeared as a promising solution, which needed to be encouraged by public policies in order to ensure its economic viability (Björk 2004; Chi Chang 2006; Asai 2020). European policy makers, whether through their member states or the European commission (EC) itself, have seized the issue by funding schemes or implementing measures to promote this practice (Lomazzi and Chartron 2014; Chartron 2016). These measures, as well as those implemented by institutions themselves (Jonchère 2013), could in part explain the development of OA publications.

Usually designed and studied under the political dimension in so far as knowledge and power are essentially linked (Koutras 2020), the practice of Open Access is a recent innovation which has however its historical basis in the notion of property, precisely in the notion of intellectual property, as coined by Locke (Koutras 2016a). Open Access appears to be a new form of ownership, which implies a new type of economic and legal relationships between publishers and authors (Koutras 2016b).

The EC had launched a pilot project within the seventh Framework Program[2], and finally published recommendations[3] in 2012 to promote open science. Member states have been encouraged to implement policies with clear and quantified objectives for the diffusion of knowledge and OA to publications from research funded by public funds. These recommendations were reinforced in 2018[4]. This new version also presents consolidated information on recent developments in open science policies of the European Union (EU) member states.

In parallel, the Open Science Policy Platform (OSPP) was launched in April 2018. The OSPP boasts, in 2020, 25 appointed members coming from universities, research organizations, finding organizations, publishers, open science platforms and libraries[5]. They meet every three to fourth months to discuss the different plans implemented within the EU, their development and their effectiveness. OSPP is also mandated to make recommendations to the EC on the political actions required to promote open science; with the ambition to make the results of European research freely accessible. This must go through progressive transition strategies from a model based on subscriptions to OA models. OSPP is complementary to initiatives driven by science actors under the supervision of EC like OpenAIRE platform (https://www.openaire.eu/), the objective of which is to promote open science and to substantially improve the dissemination and reuse of publications, research data, software and methods.

Even if it tends to harmonize under the impulsion of the EC actions, the promotion of OA over the last 30 years has been diverse between countries.

---

[2] FP7: 2007-2013, see: https://ec.europa.eu/
[3] See: https://op.europa.eu/en/publication-detail/-/publication/48558fc9-d4c8-11e1-905c-01aa75ed71a1/language-en
[4] https://op.europa.eu/en/publication-detail/-/publication/676f8a3b-62f6-11e8-ab9c-01aa75ed71a1/language-en
[5] See: https://ec.europa.eu/research/openscience/index.cfm?pg=open-science-policy-platform

The Nordic Countries (Sweden, Island, Denmark, Finland, and Norway) have early invested the field. The Norbib project, initiated in 2006 and funded by the Nordic Council of Ministers, intended to develop a common open science policy within the five northern European countries (Rabow and Hedlund 2007; OpenAIRE, Norway 2020). A few years earlier, these countries were already behind the creation of the international Directory of Open Access Journals (DOAJ) database (Björk 2019). During its first decade, DOAJ was supported by Lund University Library (Sweden).

These shared initiatives resulted from previous national actions among which the most notables were the (Bibsam Consortium 1996) in Sweden, and the DEFF (Denmark's Electronic Research Library) in Denmark. More recently, other shared initiatives involved several countries like the Alhambra Declaration on Open Access (ADOA) in May 2010, which objective is to establish common policies in Southern European Countries (Spain, Italy, Portugal, Greece, Turkey and France) (Abadal *et al.* 2010). ADOA reinforces the 2003 Berlin Declaration on Open Access to Knowledge in the Sciences and Humanities, emerged during the OA conference hosted by Max Planck Society (Germany). The Berlin declaration has an international scope with more than 250 signatory institutions (Berlin Declaration 2003).

Over the years, all the Western European countries have eventually legislated to promote OA, by amending for instance, intellectual property rules, like in Germany or in the Netherlands, in order to provide a legal framework for authors to make their research results freely available[6].

The particular situation of UK should be noticed: although the authorities did not yet implemented an open science national plan, the practice of OA is deeply rooted in the British scientific culture (whether they are publishers or institutions) and encouraged by research funders. The four UK higher education funder bodies (Research England, SFC, HEFCW and DfE) have introduced a new OA strategy for the 2021 REF (The UK Research Excellence Framework). For a publication to be eligible for the evaluation exercise, authors (or institutions) must deposit their final peer-reviewed manuscript (post-print) in a repository within 3 months of acceptance (Allen and Mehler 2019; Angelaki *et al.* 2019; Tate 2019). To help researchers and institutions to comply with the OA REF policy, several interfaces have been developed like Sherpa – RoMEO, Juliet, Fact or REF – or Open DOAR. In addition, the Registry of Open Access Repositories (ROAR) database (see: http://roar.eprints.org/), created by the University of Southampton, in 2003, lists OA repositories or archives (Bhat 2010; Pinfield *et al.* 2014; Okpala 2017).

In Eastern European countries, several initiatives have recently been developed. By the end of 2015, within the framework of the *PASTEUR4OA* project (http://www.pasteur4oa.eu/home), the Electronic Information for Libraries – EIFL (https://www.eifl.net/) and the Hungarian Academy of Sciences have organized workshop to discuss the alignment of Eastern European countries on the EC 2012 open science recommendations. The workshop made state on the multiplication of efforts by Eastern countries, especially from 2014. Estonia, Latvia, Lithuania, Poland and Slovenia have all strengthened OA communication with dedicated funds; some have even implemented open science policies and roadmaps with OA quantitative objectives[7].

---

[6] See: https://www.openaccess.nl/en/events/amendment-to-copyright-act

In view of all these efforts and initiatives, both public and private, to promote open science, it becomes imperative for political decision-makers and funders to have OA indicators to guide the measures taken. So far, the share of OA publications by institution, region or country remains the most used indicator. As it is well known, the practice of OA is fully disciplinary dependent (Gargouri *et al.* 2012; Kozak and Hartley 2013; Zhu 2017; European Commission 2020). This makes comparisons biased (Maddi 2020). Only one previous study (Archambault *et al.* 2014) have attempted to describe OA by combining spatial and disciplinary analyses. However, the study was only descriptive on the two dimensions and restricted to a short period of five years (2008-2013).

The aim of this paper is not so much to evaluate impact of public policies as to carry out a comparative study of OA practices within European countries, using data from the Web of Science (WoS) database. To measure changes over time, a systematic comparison is made between three periods 2000-2003, 2008-2011 and 2015-2018, in order to span a wide 18-years'period. The heterogeneity of practices among disciplines is controlled by the use of a normalized indicator recently developed (Normalized Open Access Indicator – NOAI (Maddi 2020)). Analysis is done both at country and regional levels, what has never been done so far. We assume that national policies may not necessarily affect all regions in the same way, while some existent studies carried out at the institutional level showed a significant disparity in terms of publications in OA worldwide (Gyawali et al. 2020; Robinson-Garcia et al. 2020). To which extend these disparities translate into variations at the regional level?

The results suggest a "correlation" between the disciplinary profile and the share of OA publications. The use of the normalized indicator highlights some particularities and provides a better framework for analysis by overcoming the disciplinary composition of countries or regions.

## Data and Methods

**Database**
The data has been extracted from the Observatoire des Sciences et Techniques' (OST) in-house database. It includes five indexes of the WoS (SCIE, SSCI, AHCI, CPCI-SSH and CPCI-S) and corresponds to the WoS content indexed through the end of March 2019. Only publications types: "article", "letter" or "review", are taken into account. In the delimitation of our perimeter, we have not made the choice to limit ourselves to publications with a DOI as is customary in studies dealing with open access topic. This choice is due to the incompleteness of the information on the DOI in the in-house database, especially for the old years (before 2010).

The WoS database offers the possibility of working on a fine, stable and validated classification (254 WoS subject categories – SC) which serves as the basis for normalization.

---

[7] For more details see: http://pasteur4oa.eu/sites/pasteur4oa/files/events/EE%20Regional%20workshops%20report%20public_long_0.pdf

**OA Status**

Since 2014, the provider of the WoS database, Clarivate Analytics, retrospectively identifies the status of OA publications. In 2017, Clarivate Analytics signed a partnership with "*Our Research*" (https://our-research.org/) that uses the Unpaywall database to identify the OA status of publications. Table 1 provides the different types of OA taken into account in the WoS database.

In this study, the OA status is considered at publication level, regardless of whether it comes out in an OA journal, that publishes only OA-type articles, or in a hybrid journal, which is fee-paying journal that gives authors the option to pay Article Processing Charges (APC) in order to publication be in OA.

Table 1: Types of Open Access provided by four different data sources.

| OA type | Unpaywall | Web of Science (database used) | Scopus | Dimensions |
|---|---|---|---|---|
| All OA | X | X | | X |
| Closed | | | | X |
| Open Access | | | X | |
| Other | | | X | |
| Bronze | X | X | | |
| Hybrid | X | | | |
| Gold | X | | | X |
|   DOAJ Gold | | X | | |
|   Other Gold | | X | | |
| Green | X | | | |
|   Green, Published | | X | | X |
|   Green, Accepted & Submitted | | | | X |
|   Green, Accepted | | X | | |

Source: Robinson-Garcia et al. (2020), page n°4.

It is important to emphasize that Table 1 shows the types "provided" in each database from a tagging point of view. In particular with regard to the Gold, Hybrid or Green statuses. For example, Unpaywall distinguishes between hybrid OA and Gold OA, while in WoS database these two types are confused in "DOAJ Gold" and "Other Gold". The same goes for the "Green" status for which the WoS distinguishes "Green, Published" and "Green, Accepted", while Unpaywall only provides the "Green" status without this distinction.

Finally, in this study we are not interested in the different OA statuses separately. Our calculations (ex. share of OA) are based on whether publications have an open access version or not, regardless of the type of access.

**Geographical/spatial attribution of publications**

Two geographical levels of analysis are proposed: the country level and the regional level, according to the NUTS classification established by Eurostat (https://ec.europa.eu/eurostat/web/nuts/background level 1, 2016 version). This nomenclature is based on administrative regions from each country, and allows comparisons between similar zones at European level.

The analysis covers the 27 countries from the EU, and 6 other European countries (Norway, Switzerland, Island, Liechtenstein, Macedonia and United Kingdom) to which the NUTS classification has been extended. The country level also includes Turkey, Kosovo, Albania and Montenegro. Eurostat provides map layers that are used as base of the spatial description.

The whole account is used to assign publications to different countries (or regions). Each publication is affected to one or more country or NUTS1 region, as far as one of its co-authors is located in the area of concern, according to the full counting method. Each publication could thus be attributed to several countries/regions. Overall, almost 1.3% of publications have one or more addresses that could not be affected to a region, varying from 0.2% to 2% among countries.

**Share of OA and Normalized Open Access Indicator (NOAI)**

The level of Open Access in each zone is approached by two indicators: the share of OA and the Normalized Open Access Indicator (NOAI).

The share of OA is the percentage of publications published in OA among all the publications; it is based on disciplinary fractional counting, the weight of a publication being inversely proportional to number of subject categories it covers. Table 2 illustrates the method of counting used with the example of a publication co-signed by two authors affiliated with two institutions, one in France and the other in Germany. This publication is also assigned to four different WoS categories and two disciplines.

**Table 2: Counting method for multidisciplinary publications**

| Field fraction | Discipline fraction | Country fraction (whole count) | |
| --- | --- | --- | --- |
| | | France:1 | Germany: 1 |
| Medical Informatics: **0.25** | Computer science: **0.5** | 0.25*1 = 0.25 | 0.25*1 = 0.25 |
| Computer Science, Information Systems: **0.25** | | 0.25*1 = 0.25 | 0.25*1 = 0.25 |
| Public, Environmental & Occupational Health: **0.25** | Medical research: **0.5** | 0.25*1 = 0.25 | 0.25*1 = 0.25 |
| Primary Health Care: **0.25** | | 0.25*1 = 0.25 | 0.25*1 = 0.25 |

While it also possible to consider a fractional counting on the geographical dimension, and a combined fractional counting on both dimensions. We apply the disciplinary fractional counting, as we only focus on disciplinary contribution to OA.

The Normalized Open Access Indicator (NOAI) is a relative index that allows comparisons while controlling for disciplinary specialization of countries or regions. The indicator is constructed by analogy to the well-known normalization methods practiced in bibliometrics (see: Waltman 2016). NOAI is calculated in two steps:

- First, an OA index is calculated for each subject category ($OAI_{SC}$) by dividing the share of OA within the zone ($NOA_{SC} / N_{SC}$) by the share of OA worldwide for the SC ($NOA(W)_{SC} / N(W)_{SC}$):

$$OAI_{SC} = \frac{NOA_{SC} / N_{SC}}{NOA(W)_{SC} / N(W)_{SC}}$$

- Second, the NOAI is calculated as the average of OA indexes for each SC, weighted by the number of publications within the SC:

$$NOAI = \frac{\sum (OAI_{SC} \times N_{SC})}{\sum N_{SC}}$$

The NOAI can be interpreted as the ratio of OA practice for an object of interest (here, the country or region) regarding to a reference. It is interpreted relatively to the value 1.

**Spatial representations and discretization**

These indicators are represented in choropleths maps for each of the three periods of study ([2000-2003], [2008-2011] and [2015-2018]). We choose to apply a unique discretization based on the overall distribution of indicators, in order to allow the comparison between periods.

The share of OA is represented in 9 classes, obtained by using the Jenks natural breaks optimization method (see: Jenks, 1967). The gradation of colors indicates how high the indicator is.

The NOAI is represented in 7 classes. As its interpretation is relative to 1, we retain a manual symmetric classification with a central "neutral" class and two different colors in order to put in evidence the contrast between zones under and over than 1.

## Results

This section is organized as follows. First, we develop a comparative analysis of countries in terms of the openness of scientific publications using the share of OA publications. This analysis is followed by a characterization of the disciplinary profile of countries (or regions) which could partly explain the disparities noted on the share of OA publications. Finally, we make a comparison using the NOAI which allows overcoming the differences due to disciplinary practices and the scientific profile of countries or regions.

**Comparison between countries and regions on OA indicators**

Figure 1 shows a strong increase in the share of OA publications in all European countries over the three periods: 2000-03, 2008-11 and 2015-18. However, there are great discrepancies between countries.

Figure 1: OA share by country, 2000-03, 2008-11 and 2015-18

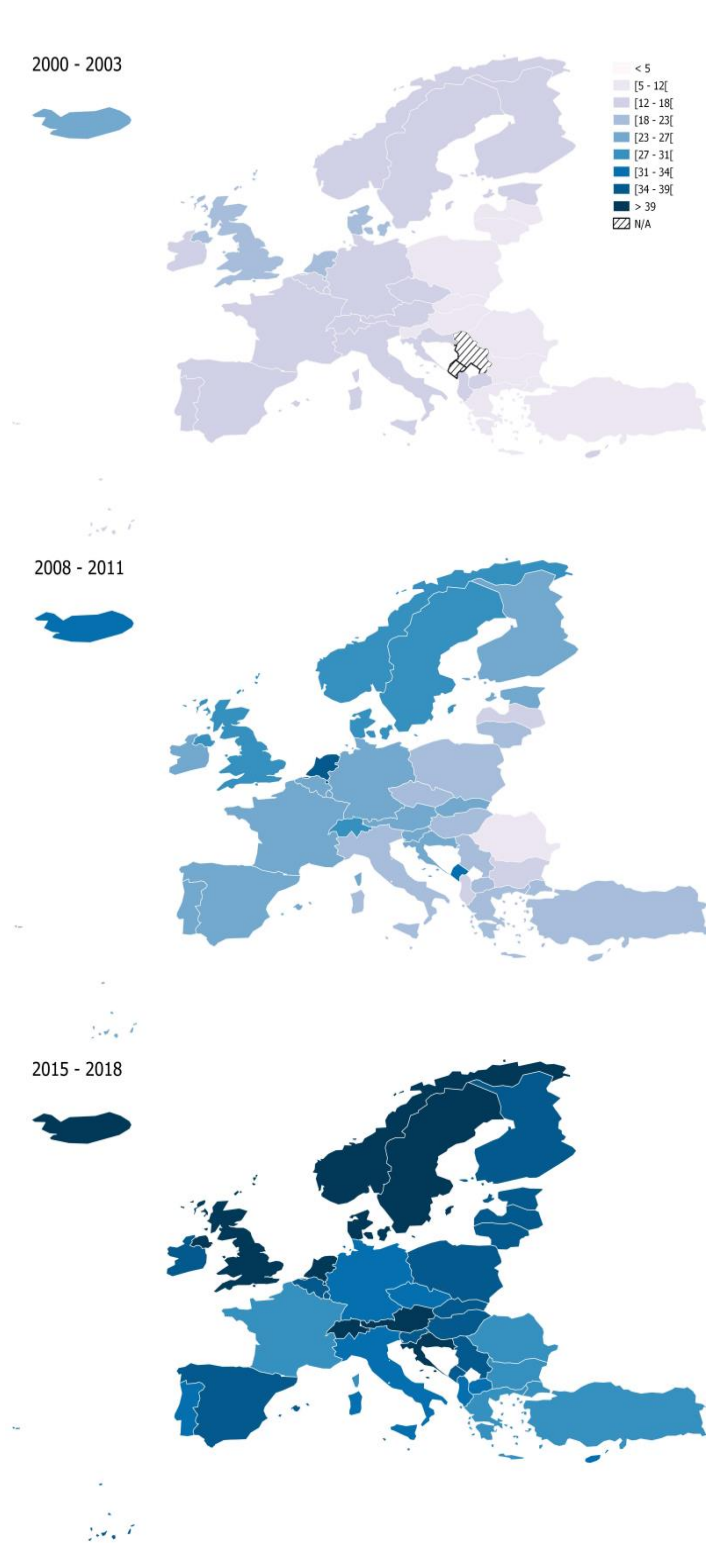

Iceland, UK, Denmark and the Netherlands had the highest shares of OA publications during the first period (2000-03). Their shares vary between 20 and 27% while in the rest of European countries; shares did not exceed 18%. These countries remain among the most open during the last period 2015-18 with shares that exceed 39%.

In addition, some countries have made a huge jump in the share of open access publications. This is particularly the case of Sweden, Norway, Switzerland, the Czech Republic and Croatia which now have shares exceeding 39%. Other countries have also made considerable progress, but to a lesser extent with rates varying between 31 and 39%. This is the case of several Eastern European countries, as well as Spain and Portugal.

Among the European countries, countries with the lowest progression in OA share, we find France, Romania, Bulgaria, Greece and Turkey, with a share that varies between 27 and 31%.

Furthermore, it is important to note that countries which have made the most progress are generally countries which have implemented early OA policies, like the northern European countries

Figure 2: OA share by NUTS1, 2000-03, 2008-11 and 2015-18

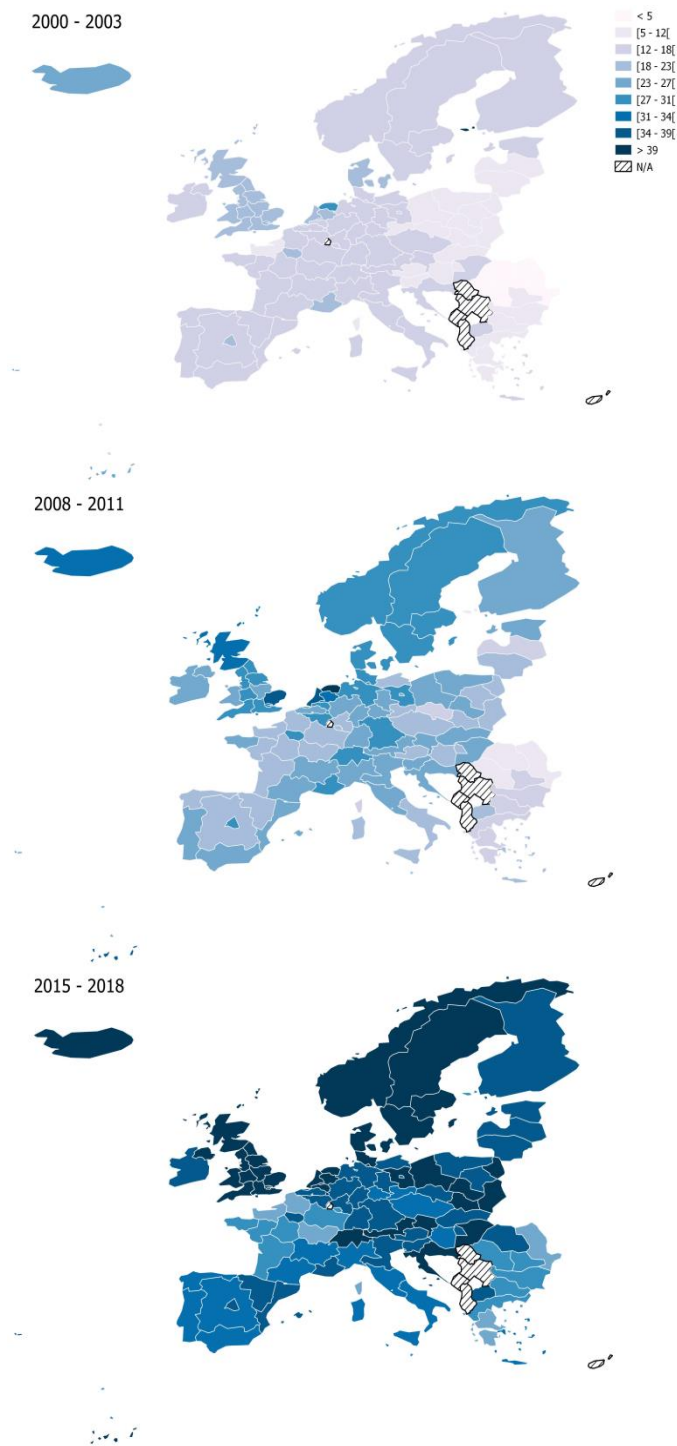

From figure 2 we can see that, overall, the share of OA publications is similarly distributed within the regions. However, some peculiarities should be noted. Among European regions, the UK regions are the most open, with OA shares sharply increasing. "Yorkshire-and-The-Humber" and "East-of-England" are the regions with the highest growth of OA publications, shares rising respectively from 19 and 22% in 2000-03 to 54 and 57% in 2015-18. This can be explained in particular by the fact that in these regions there is a very strong incentive to disseminate research results from institutions. In addition to the OA requirements of the REF, several universities have adopted their own OA policies with dedicated staff and funds. For example, in Yorkshire-and-The-Humber, the Leeds Beckett University researchers should add the bibliographic data relating to their outputs to their Symplectic Elements account (which is research management system that collects all research activity of the university). Likewise, Sheffield Hallam University (SHU) Board endorsed an OA policy in 2014 and it was updated in 2019. Indeed, authors must make available a copy of scientific production resulting from SHU research. Another example from the University of Leeds which establishes OA policy requires that all newly accepted publications should recorded in Symplectic (the University publications database) within 3 months of acceptance.

In the East-of-England region, we can cite The University of Cambridge case. The University of Cambridge has set up an "Open Access Service based in the Office of Scholarly Communication". The Open Access Service reduces the administrative burden of OA policies on academics and makes as many outputs OA as possible in accordance with copyright and license agreements. Open Access Service also assists researchers in depositing their publications into the institutional repository.

Apart from the UK, several regions have made significant progress in terms of OA. This is particularly the case for regions Noord-Nederland, West-Nederland and Oost-Nederland in the Netherlands with rates of 50, 47 and 46% respectively in 2015-18 (against 28, 23 and 18% in 2000-03). This is also the case for the regions Östra-Sverige in Sweden and Brandenburg in Germany with a rate of 43% in 2015-18 (against 18 and 16% respectively).

**Open Access and disciplinary differences**

In this section, we first analyze the practices of OA by discipline. Secondly, the disciplinary profile of countries of the study. The aim is to see to what extent the differences observed in figures 1 and 2 can be explained by disciplinary specificities.

Figure 3: Share of OA publications by discipline in the WoS database

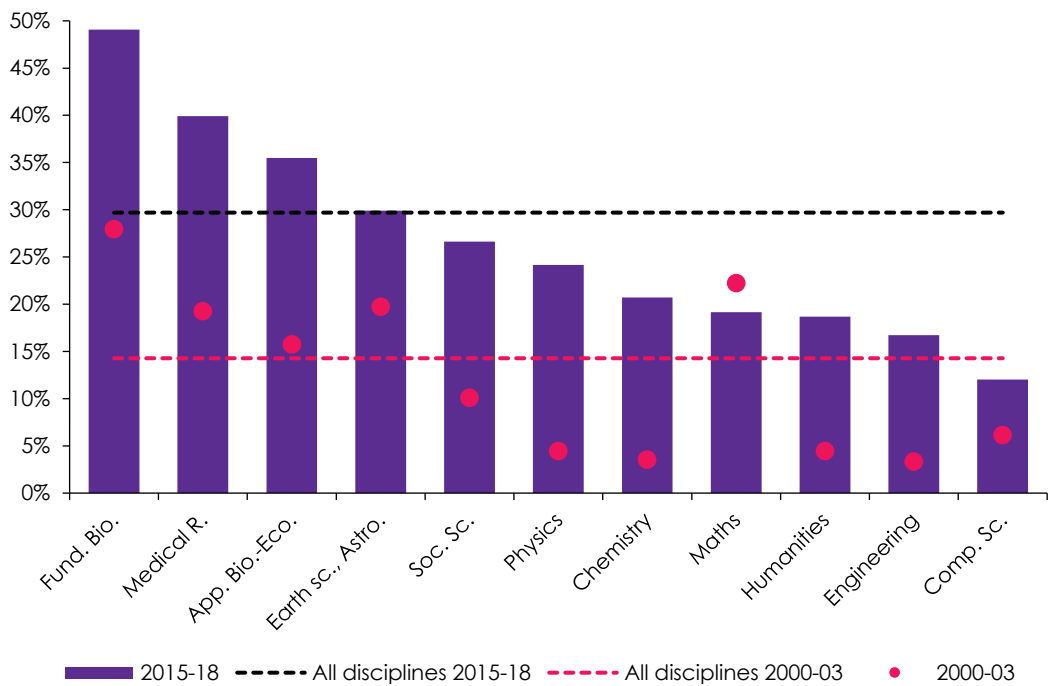

Figure 3 shows that the global share of OA publications has roughly doubled between the two periods 2000-03 and 2015-18. It went from 15 to 30%. Nevertheless, there are significant disparities between disciplines. The share is very high in fundamental biology (50% in 2015-18), medical research (40%) and applied biology - ecology (37%). Conversely, the share is significantly lower in computer science (12%) and engineering (17%). In addition, it is important to note the strong progression over the two periods of almost all of the disciplines. The only discipline in which the share of open access publications has decreased is mathematics. The explanation for this decline requires further study, which is not the aim of this article.

The differences between countries in terms of the openness of scientific publications observed in Figures 1 and 2 could therefore be linked to their disciplinary profile. Thus, the more scientific production in a country is focused on natural sciences, the greater is the share of OA publications. Therefore, we calculated the specialization indexes[8] for the countries of study presented in Table 3.

Table 3: Specialization indexes by county (2015-18, fractional counting).

| country | Fund. Bio. | Medical R. | App. Bio.Eco. | Earth sc.Astro. | Soc. Sc. | Physics | Chemistry | Maths | Humanities | Engineering | Comp. Sc. |
|---|---|---|---|---|---|---|---|---|---|---|---|
| **Austria** | 1,09 | 1,07 | 1,04 | 1,00 | 0,92 | 0,87 | 0,77 | 1,54 | 1,18 | 0,88 | 1,26 |
| **Belgium** | 1,11 | 1,12 | 1,18 | 0,87 | 1,27 | 0,81 | 0,7 | 0,88 | 1,94 | 0,82 | 0,80 |
| **Bulgaria** | 0,61 | 0,48 | 1,62 | 1,02 | 0,71 | 1,41 | 1,53 | 2,35 | 0,58 | 0,87 | 0,92 |
| **Croatia** | 0,55 | 0,91 | 1,30 | 1,07 | 1,99 | 0,45 | 0,7 | 1,54 | 1,99 | 1,11 | 0,60 |
| **Cyprus** | 0,60 | 0,65 | 0,72 | 1,09 | 2,31 | 0,61 | 0,45 | 1,10 | 2,22 | 1,35 | 1,71 |
| **Czech** | 0,98 | 0,50 | 1,39 | 1,09 | 1,34 | 1,10 | 1,16 | 1,44 | 1,19 | 1,00 | 1,07 |
| **Denmark** | 1,16 | 1,48 | 1,12 | 0,95 | 1,37 | 0,6 | 0,49 | 0,48 | 1,04 | 0,8 | 0,65 |
| **Estonia** | 0,94 | 0,41 | 1,41 | 1,67 | 1,48 | 0,85 | 0,78 | 0,76 | 3,18 | 0,87 | 1,05 |
| **Finland** | 0,95 | 0,97 | 1,11 | 1,12 | 1,71 | 0,72 | 0,61 | 0,82 | 1,33 | 0,94 | 1,27 |
| **France** | 1,10 | 1,08 | 0,8 | 1,17 | 0,63 | 1,16 | 0,86 | 1,66 | 1,04 | 0,90 | 1,11 |
| **Germany** | 1,19 | 1,05 | 0,84 | 0,97 | 0,90 | 1,21 | 0,94 | 1,03 | 1,17 | 0,88 | 0,91 |
| **Greece** | 0,66 | 1,21 | 0,92 | 1,44 | 0,95 | 0,69 | 0,62 | 1,06 | 0,72 | 1,11 | 1,61 |
| **Hungary** | 1,21 | 0,91 | 1,4 | 1,00 | 0,79 | 0,89 | 0,93 | 2,22 | 0,99 | 0,83 | 0,92 |
| **Iceland** | 0,79 | 1,03 | 1,11 | 2,19 | 1,77 | 0,65 | 0,31 | 0,47 | 1,78 | 0,79 | 0,97 |
| **Ireland** | 0,98 | 1,17 | 1,23 | 0,73 | 1,71 | 0,64 | 0,67 | 0,62 | 1,65 | 0,75 | 1,09 |
| **Italy** | 1,03 | 1,32 | 0,96 | 1,13 | 0,73 | 0,91 | 0,67 | 1,16 | 0,99 | 0,95 | 0,95 |
| **Latvia** | 0,35 | 0,29 | 1,34 | 0,86 | 2,85 | 0,86 | 1,17 | 0,55 | 0,92 | 1,41 | 1,13 |
| **Liechtenstein** | 0,38 | 0,72 | 0,31 | 0,1 | 3,25 | 0,82 | 1,69 | 0,00 | 1,32 | 0,88 | 1,30 |
| **Lithuania** | 0,6 | 0,54 | 1,21 | 0,86 | 1,91 | 1,10 | 0,96 | 1,31 | 2,17 | 1,22 | 0,61 |
| **Luxembourg** | 0,97 | 0,48 | 0,58 | 0,80 | 1,91 | 0,85 | 0,51 | 2,01 | 1,17 | 1,08 | 3,37 |
| **Malta** | 0,51 | 1,16 | 0,88 | 0,94 | 1,59 | 0,61 | 0,39 | 1,03 | 2,17 | 0,97 | 1,92 |
| **Netherlands** | 1,10 | 1,57 | 0,80 | 0,91 | 1,71 | 0,54 | 0,43 | 0,47 | 1,63 | 0,63 | 0,71 |
| **Norway** | 0,84 | 1,1 | 1,08 | 1,6 | 2,01 | 0,39 | 0,36 | 0,74 | 1,62 | 0,98 | 0,84 |
| **Poland** | 0,82 | 0,74 | 1,4 | 1,26 | 0,65 | 1,18 | 1,16 | 1,47 | 0,73 | 1,09 | 1,09 |
| **Portugal** | 0,94 | 0,79 | 1,17 | 1,32 | 1,18 | 0,66 | 0,85 | 1,13 | 1,07 | 1,08 | 1,64 |
| **Rep. of North Macedonia** | 0,46 | 0,66 | 0,92 | 1,18 | 2,23 | 0,78 | 0,62 | 1,41 | 0,83 | 1,02 | 2,44 |
| **Romania** | 0,45 | 0,45 | 0,57 | 1,04 | 1,87 | 0,95 | 1,37 | 2,00 | 0,92 | 1,46 | 1,13 |
| **Slovakia** | 0,84 | 0,4 | 1,09 | 1,17 | 1,96 | 1,04 | 1,01 | 1,09 | 1,6 | 1,2 | 0,99 |
| **Slovenia** | 0,72 | 0,67 | 1,15 | 0,91 | 1,26 | 0,83 | 1,14 | 2,07 | 2,1 | 1,03 | 0,97 |
| **Spain** | 0,95 | 0,99 | 1,24 | 1,12 | 1,26 | 0,69 | 0,84 | 1,08 | 1,68 | 0,81 | 1,09 |
| **Sweden** | 1,06 | 1,23 | 0,84 | 0,98 | 1,66 | 0,69 | 0,64 | 0,68 | 1,17 | 0,89 | 0,91 |
| **Switzerland** | 1,28 | 1,18 | 0,95 | 1,15 | 0,99 | 1,08 | 0,77 | 0,87 | 1,16 | 0,71 | 0,79 |
| **Turkey** | 0,54 | 1,53 | 1,14 | 0,91 | 0,91 | 0,76 | 0,81 | 1,22 | 0,65 | 0,94 | 0,78 |
| **UK** | 1,07 | 1,14 | 0,77 | 0,99 | 1,88 | 0,72 | 0,56 | 0,74 | 2,37 | 0,71 | 0,82 |

Table 3 shows the specialization indexes by country for the period 2015-18. The specialization index is calculated for an actor (country, region, institution, etc.) in a given discipline. It is defined by the share of discipline in the actor's publications, related to the same share at the world level. The specialization index is interpreted in relation to the value 1; the more it is greater than 1 the more the actor is specialized in the discipline. The results in Table 1 confirm the hypothesis that countries specializing in at least one of the three disciplines (fundamental biology, medical research and applied biology - ecology) have the highest share of OA, with a few exceptions. Thus, over the 2015-18 period, all countries with a share of OA publications greater than that of the world (which is 30%: see Figure 3) are at least specialized in one of the three disciplines with the highest share of OA.

---

[8] (Rousseau 2018) proposed an alternative indicator, the F-measure. Our calculations showed that the Pearson correlation between the traditional specialization index and F-measure is highly significant, at 0.74, with p <0.001. (Mescheba et al. 2019) also show that F-measure is also open to criticism for its dependence on the volume of publications. Otherwise, regardless of the disciplinary indicator used to measure specialization, disciplinary disparities will remain and normalization (NOAI) remains important in all cases.

Thus, within the Nordic countries Denmark specializes in the three disciplines and especially in medical research with a specialization index 48% higher than the world average. Sweden, in addition to social sciences and humanities, is also specialized in medical research and to a lesser extent in fundamental biology. Norway has specialization indexes around the world average in medical research and applied biology - ecology. Norway is more specialized in social sciences and humanities and Earth sciences—Astronomy—Astrophysics and is not specialized in fundamental biology. Although Norway has a disciplinary profile similar to that of Iceland, the Iceland's share of OA publications is higher than that of Norway (see Figure 1).

The other countries with a high share of OA publications, such as the UK, the Netherlands and Switzerland are both specialists in fundamental biology and medical research. In contrast, there are countries like Greece that specialize in computer science and do not specialize in fundamental biology or applied biology - ecology. However, Greece is specialized in medical research.

Eastern European countries are generally specialized in applied biology - ecology and are not specialized in medical research or fundamental biology. Their share of OA is generally lower than that of the Nordic countries, but remains higher than that of the world.

The comparison between disciplinary structure (Table 1) and share of OA publications (Figures 1 and 2) suggests that there is a link between the two. Countries with a strong science (biology and medical research) component generally have higher shares of OA publications. Consequently, analyzing the simple share of OA publications does not allow a distinction to be made between the part which is due to the disciplinary profile and that which results from other factors such as open science public policies and the involvement of researchers and institutions in the OA movement. The share of OA publications does not seem to us a good indicator for making comparisons between actors with different profiles.

Constructing a field normalized indicator allows to correct OA share of the disciplinary composition of countries. Although the normalized indicator does not allow a distinction to be made between what relates to public policies and what is due to the involvement of national actors in the OA movement, it nevertheless makes it possible to overcome disciplinary specificities.

Figure 4 shows evolution of the NOAI by country. NOAI compares the share of OA of countries to that of the world. The presentation in three periods makes it possible to analyze evolution of this relationship between the national (or regional) share and the world share. We note that the majority of European countries have progressed more than the world over time. Whereas in the first period (2000-03), more than half of the countries were below the world average, during the last period (2015-18), the NOAI is higher than the world average for most countries. However, some countries remain around (even lower) the world average, like Germany, France, Italy, Greece and Turkey with an indicator between 0.9 and 1.1.

Figure 4: NOAI by country, 2000-03, 2008-11 and 2015-18

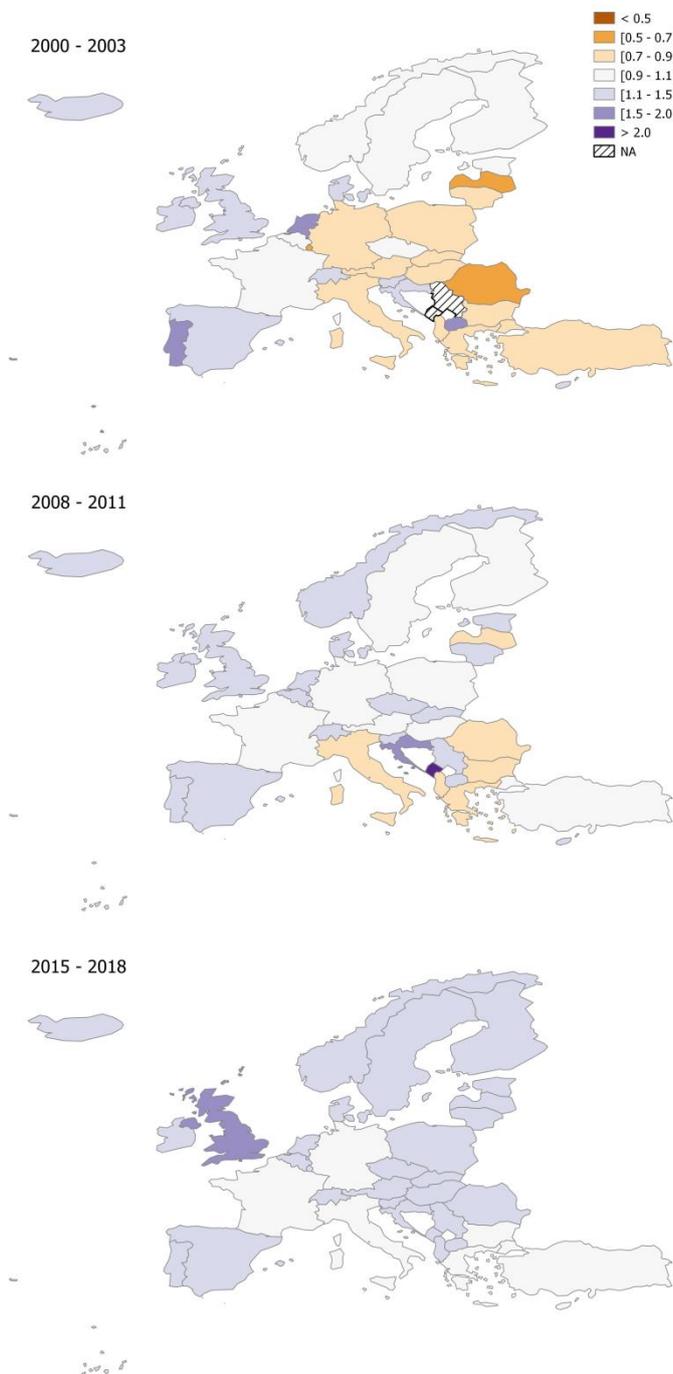

In 2015-18, the majority of countries have an indicator 10 to 50% higher than the world average, with the exception of the UK whose indicator is greater than 50% (NOAI between 1.5 and 2). Otherwise, some countries have made more progress than others. This is particularly the case for Latvia and Lithuania, which move respectively from an indicator 31% and 26% below the world average to an indicator 48% and 46% higher. Poland has also made significant progress from a 22% lower indicator than word average to an indicator 38% higher. In addition, it is important to note that countries with low NOAI are progressing but to a lesser extent. Conversely, other countries have seen their relative OA share (NOAI) decline over time. This is particularly the case of Portugal which goes from a NOAI from 1.95 to 1.10. The same goes for Ireland, going from 1.40 to 1.12. This relative decline was not observable with the gross rate of open access.

The comparison between figures 1 (share of OA) and 4 (NOAI) highlights several observations. During the 2000-03 period, it can be seen that many countries have similar OA shares. The normalized indicator (Figure 4) shows a great contrast between countries. The countries of Eastern Europe have the lowest indicators compared to the world average. Portugal had a normalized indicator similar to that of the Netherlands, while the share of OA is higher for the Netherlands. Over the period 2015-18, we observe that the countries of Eastern Europe have lower shares of OA than those of Finland or Sweden for example. After normalization, in Figure 4, we see that they have similar indicators (same class). This is partly explained by the fact that the Eastern European countries are specialized (see Table 1 and Figure 3) in disciplines with low OA shares in the WoS database. In addition, Figure 4 shows that the UK is the most open country, while in Figure 1 we see that it is in the same class as a number of other countries like Finland, Norway, Switzerland and the Czech Republic.

The analysis in terms of evolution highlights important lessons. In Figure 1, the reader can note only the high OA shares of the Nordic countries and some central European countries, while by analyzing the normalized indicator; we can better observe the enormous progress made by the Eastern European countries.

Figure 5: NOAI by NUTS1, 2000-03, 2008-11 and 2015-18

Figure 5 does not reveal specific trends at the regional level, which tends to suggest that public policies carried out at the national level are the major factor in stimulating OA. Nevertheless, a discussion around the particularities of France would possibly merit a meso or micro analysis (regional or institutional factor) given that the indicators are not similarly distributed over the whole of the territory (over the last period).

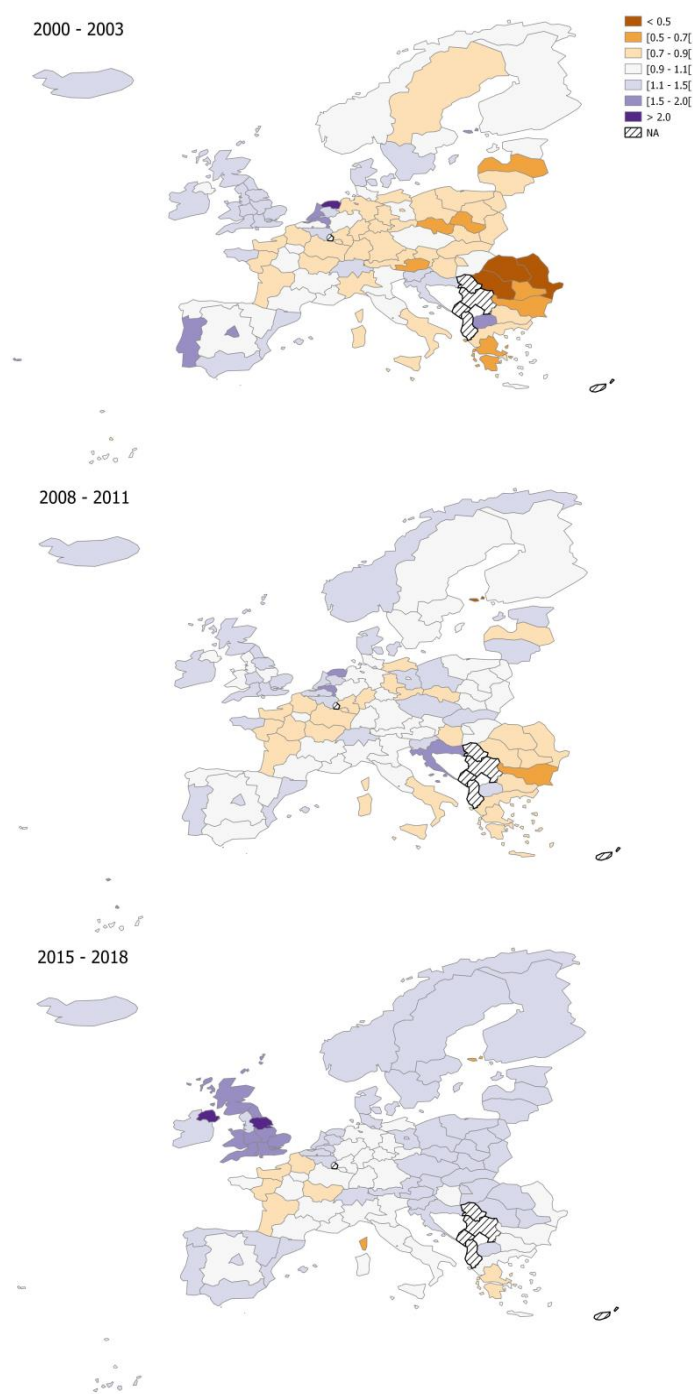

If we look in details, we observe that the majority of European regions passed above the world average in the last period (2015-18). Some regions have made spectacular progress in terms of OA publications, such as, in northern regions, Romania, which goes from an indicator 50% lower than the world average (in 2000-03), to an indicator 10 to 15% higher (in 2015-18). This is also the case in several regions of Eastern Europe countries.

Other regions made less progress than the world, when they had indicators twice the world average in 2000-03. This is particularly the case for "Groningen", "Northern-Holland" and "Over-Ijssel" in Netherlands, "Madrid" in Spain and "Centro", "Lisboan", "Alentejo" and "Algarve" in Portugal, which changes to a NOAI between 1.1 and 1.5 (against more than 1.5 in 2000-03).

In addition, certain regions, notably in France and in Greece, remained below the world average throughout the three periods. In France, this is the case for "Hauts-de-France",

"Normandie", "Pays-de-la-Loire", "Nouvelle-Aquitaine", "Bourgogne-Franche-Comté" and "Corse" with an indicator between 0.5 and 0.9 (10 to 50% lower than the world average). This is also the case of regions of "The Peloponnese", "Central Greece" and "Thessaly" in Greece.

Across all regions, results highlight that the UK regions have the highest indicators. In particular: "Northern Ireland" and "Yorkshire and the Humber" with a normalized indicator greater than 2. Almost all of the other regions of the UK have indicators between 1.5 and 2. These performances in terms of OA are probably strongly linked to the policy applied within the research evaluation framework (REF) which mandates researchers to make their scientific production freely available. A dedicated study is necessary to be able to confirm this. Likewise, for a more detailed interpretation of regional particularities requires further study.

## On the importance of OA types and data sources to monitor OA uptake

When it comes to analyzing open access publications, the choice of database can affect the results obtained. The objective of this section is to provide an overview on the different types of OA and their indexing in the databases. This allows to underline that the results obtained in this article are inherent to the database used; i.e. the WoS.

The modalities by which readers have open access to scientific publications are diverse and depend on a combination of editorial and dissemination practices. These different aspects are reviewed insofar as they have an impact on the indexation of open access publications in the databases.

The two main types of open access are, on the one hand, the golden route (Gold OA), and its variants (platinum, etc.), which is defined by the mode of publication and, on the other hand, the green route (Green OA) which is defined by the broadcast mode. Their indexation in databases is different.

This typology of access routes emphasizes that it is important to distinguish the status of the publication from that of the journal. A journal can have three statuses: subscription-based (closed), open access or hybrid. An open access journal publishes all of its articles in open access: the articles are in open access through the golden channel. The journal's remuneration is based on publication fees that it charges (Article Processing Charges - APC) and / or grants from a public or private institution. A hybrid journal is a subscription journal which allows authors to make their article available to readers free of charge, on payment of a publication fee (APC). The hybrid journal thus publishes both articles accessible to readers, for a subscription or an access fee, and open access articles. An open article from a hybrid journal has a golden OA status in the WoS database.

In the golden road to open access, the articles are published under a Creative Commons (CC) type license in journals listed in particular in the Directory of Open Access Journals (DOAJ); they are open access as soon as they are published. In December 2020, the DOAJ had 15,600 journals. Although the selection of DOAJ journals is based on criteria of academic quality (presence of peer review policy, stability over time, etc.) and the quality of metadata, the indexation rate of these journals remains low in international databases (like WoS or Scopus).

For example, only 12% of DOAJ journals in information science are indexed in Scopus (Sahoo et al. 2017). However, the indexing of OA journals is progressing rapidly in international databases (Björk, 2019). The rate of OA journals in the JCR (Journal of Citations Report) of the WoS database was 2.6% in 2003 (McVeigh, 2004), while the rate is now in the order of 18% in WoS and Scopus (Björk, 2019).

There are two reasons for the increase in the number of OA journals indexed in international databases. First, the number of journals which are born in OA has increased significantly over the past 15 years (Laakso and Björk, 2012; Björk, 2019). Second, many journals that are already indexed convert to OA. Thus, 53% of OA journals indexed in Scopus were subscription-based before converting to 100% OA (Solomon et al. 2013).

Publications can have a so-called "bronze" open access status, which includes articles published in journals that do not apply a CC license or that have an unidentified open access license in databases. The bronze type can also be due to the fact that the status is temporary. For example, an article available as read-only on the publisher's site will have an OA bronze status: after the embargo period, the author can make the final version of the article available in an open archive. It therefore becomes a publication in green OA. The bronze status is therefore not stable and its identification in the databases depends on the access dates. The status of freemium publications is difficult to trace in databases and can be akin to the bronze path. In the so-called "freemium" diffusion model, the publisher makes all or part of a publication available in a simple format (html or text for example), then it gets remuneration by giving access to formats more convenient for the reader.

The green OA concerns articles deposited in an open archive. Journals that are not in open access may allow authors to deposit their future publication in an open archive - either the version before peer review (preprint) or the version accepted for publication (postprint). The authorization often comes after an embargo period of up to two years or more. When depositing the publication in the archive, the SHERPA / RoMEO interface allows authors to know the copyright policies and restrictions per journal. Journals can also dump their entire content into an archive like PubMed Central. Therefore, an article can have several OA statuses at the same time; for example, an article can be published in an open access journal (golden route), then be deposited in an open archive (green route).

Björk et al. (2014) provide a detailed analysis on the green OA. They define four types of scientific productions in green OA: working papers, submitted manuscripts, manuscripts accepted for publication and published articles. Full texts can be found in three main sources: institutional archives, thematic archives and authors' personal / institutional web pages. In terms of indexing in databases such as WoS or Scopus, several studies show that the rate of publications in green OA is relatively low - around 12% (Hajjem et al. 2005; Björk et al. 2010 ; Gargouri et al. 2012; Laakso et al. 2012; Björk et al. 2014). Piwowar et al. (2018) publication compare three samples of 100,000 articles with DOI extracted from three databases: Crossref, WoS and Unpaywall. They estimate that the rate of green OA posts varies between 11% and 12% in WoS. The estimated rate is much lower in Crossref; it is in the range of 4.5 to 5%. In Unpaywall the green OA rate is estimated to be between 8.8 and 9.4%.

The sustainability of the green OA status depends on the one hand on respect for copyright and on the other hand on the sustainability of the platform where the articles are deposited.

Some authors post the publisher's version in open archives when in many cases this is not allowed. Once the publisher becomes aware of the deposit, the publications concerned will be removed from the archive or possibly replaced by an authorized version (preprint). The lifespan of certain archives can thus be limited and therefore the duration of access to the articles in green OA which are there. This is why it is preferable to favor publication in a perennial archive (ArXiv type) which implements a control of the deposit of the document (see SHERPA/RoMEO).

This overview highlights that the reliability and availability of OA information varies according to the respective status of journals and publications. While the information on gold open access publications can be considered reliable and stable, the same cannot be said for bronze, which is temporary. As open archives are continuously fed by authors or journals, information on green open access is also evolving. Indexing in international databases strongly depends on these elements.

Discussion and conclusion

Through this article, we have mapped and analyzed the OA publications produced by the European countries over three periods (2000-03, 2008-11 and 2015-18). We calculated two OA indicators from WoS data: the share of OA publications and the NOAI.

Results show that the share of OA has increased significantly for all European countries. However, the analysis of the relative evolution compared to the world (NOAI) shows important disparities between countries and highlights some regional particularities.

At country level, we can identify three country profiles with the highest OA indicators. First, the countries that "mandate" researchers to deposit their scientific production in OA. This is particularly the case of the UK, where one of the eligibility conditions for the REF is that the publication be in OA. Second, countries like the Nordic countries that have implemented early OA policies (since the 1990s). These countries have seen their share of OA increase significantly over time. The same is true for the normalized indicator (NOAI). Finally, there are countries whose institutions are widely involved in the OA movement. This is particularly the case in the Netherlands, UK and certain Eastern European countries. In these countries, the institutions communicate a great deal on the issue of access to research results and support researchers to make their publications available on open archives. Some institutions even have funds and staff dedicated to open science.

We can also note the importance of developing institutional archives that facilitate the opening of publications. Several countries are ahead of this question, like Norway with more than 40 repositories and archives and the UK where a good part of universities have their own archives/repositories. Other countries are in "catch-up mode" whose policies on open science are recent with a lack of archives or repositories at national and institutional level. In these countries, until a few years ago, researchers were not strongly encouraged to make their research results available. The issue of the impact of publications was superior to that of OA (although the two were not contradictory). Therefore, there was a lack of support for researchers and institutions in the area of open science. This is compounded by the fact that many of these countries do not have a highly developed publishing industry. Consequently, they have a weak negotiating power with editors as regards the agreements around the "big deal" to pass gradually to a completely OA economic model.

At regional level, there is no particular lesson to be learned from it. The distribution of open access publications within countries is uniform. Nevertheless, in some cases (ex. France), we observed some peculiarities and differences at regional level. Explaining these differences requires further analysis at the institutional and journal levels. For instance, several studies of regional OA journals indicate that the rate of inclusion OA journals in the DOAJ database varies a lot (Björk 2017; Björk 2019). Björk (2017) showed that only 18% of 6,509 OA journals hosted in 15 regional portals are indexed in DOAJ. This will be the subject of our next studies on WoS database. It would also be interesting to analyze, by country and region, the question of journals which move from a subscription based economic model to an "Articles Processing Charges" based model (OA). (Solomon et al. 2013) for example showed that 79% of OA journals from globally dominating publishing countries (USA, the Netherlands, UK and Germany) were born OA, but this rate was only 32% for other countries.

Finally, despite the fact that the link between "open science public policies" and OA publications seems obvious, it is essential to perform empirical models to demonstrate this. Several factors can indeed influence OA indicators; especially databases coverage. Hence, the data source used is essential for estimating the share of open access publications. Not all research outputs are published in scientific journals. For example, within the ArXiv repository, many documents have not reached the publication stage in journals listed in WoS or Scopus. Including working papers or unsubmitted articles deposited in open archives into the study can greatly increase the estimated rate of open access publications.

Physics and mathematics are the disciplines with the highest share of open access publications, the majority of which are green OA. In contrast, a recent study that uses the WoS database showed that these disciplines have the lowest share of OA (Maddi, 2020). This difference is explained by the fact that in international databases, only articles published in scientific journals covered are taken into account. For example, an article in physics deposited in an open archive will only be taken into account if this article has also been published in journal indexed by the databases in question.

The normalized indicator used in this article allows a correction for this bias. Thus, even if the share of OA publications is largely underestimated in WoS, the fact of relating the country's OA share to that of WoS gives a better idea of the degree of openness of its research. This, of course, does not solve all WoS coverage issues.

# References

Abadal, E. (Ed ), Anglada, L. (Ed ), Abad, F., André, F., Badolato, A.-M., Barthet, E., … Weil-Miko, C. (2010). *Open Access in Southern European Countries*. Madrid : Fundacion española para la ciencia y la tecnología, 2010.

Allen, C. and Mehler, D.M.A. (2019). Open science challenges, benefits and tips in early career and beyond. *PLOS Biology* **17**(5):e3000246. doi: https://doi.org/10.1371/journal.pbio.3000246.

Angelaki, M., Garnier, M., Gini, B., Grimstone, P., Hansen, D., Sartori, A., … Varela Fuentes, E. (2019). REF 2021 open access requirements: a practical introduction to uploading publications. doi: https://doi.org/10.17863/CAM.40669.

Archambault, É., Amyot, D., Deschamps, P., Nicol, A., Provencher, F., Rebout, L. and Roberge, G. (2014). Proportion of Open Access Papers Published in Peer-Reviewed Journals at the European and World Levels—1996–2013. *Copyright, Fair Use, Scholarly Communication, etc.*

Asai, S. (2020). Market power of publishers in setting article processing charges for open access journals. *Scientometrics*. doi: https://doi.org/10.1007/s11192-020-03402-y.

Berlin Declaration (2003). *Berlin Declaration*. Available at: https://openaccess.mpg.de/Berlin-Declaration [Accessed: 19 March 2020].

Bhat, M.H. (2010). Open access repositories: a review. *Library Philosophy and Practice*.

Bibsam Consortium (1996). *Bibsam Consortium*. Available at: https://www.kb.se/samverkan-och-utveckling/oppen-tillgang-och-bibsamkonsortiet/open-access-and-bibsam-consortium/bibsam-consortium.html [Accessed: 17 March 2020].

Björk, B.-C. (2004). Open Access to Scientific Publications – An Analysis of the Barriers to Change?

Björk, B.-C. (2014). Open access subject repositories: An overview. Journal of the Association for Information Science and Technology, 65(4), 698–706.

Björk, B.-C. (2017). Journal portals – an important infrastructure for non-commercial scholarly open access publishing. *Online Information Review* **41**(5):643–654. doi: https://doi.org/10.1108/OIR-03-2016-0088.

Björk, B.-C. (2019). Open access journal publishing in the Nordic countries. *Learned Publishing* **32**(3):227–236. doi: https://doi.org/10.1002/leap.1231.

Bruns, A., Rimmert, C. and Taubert, N. (2020). Who pays? Comparing cost sharing models for a Gold Open Access publication environment. *arXiv:2002.12092 [cs]*.

Chartron, G. (2016). Stratégie, politique et reformulation de l'open access. *Revue française des sciences de l'information et de la communication*(8). doi: https://doi.org/10.4000/rfsic.1836.

Chi Chang, C. (2006). Business models for open access journals publishing. *Online Information Review* **30**(6):699–713. doi: https://doi.org/10.1108/14684520610716171.

European Commission (2020). *Open Access | Open Science - Research and Innovation - European Commission*. Available at: https://ec.europa.eu/research/openscience/index.cfm?pg=openaccess# [Accessed: 9 March 2020].

Gargouri, Y., Larivière, V., Gingras, Y., Carr, L., & Harnad, S. (2012). Green and Gold Open Access Percentages and Growth, by Discipline. In, 17th International Conference on Science and Technology Indicators (STI), Montreal, CA, 05 – 08 Sep 2012. 11pp, http://eprints.soton.ac.uk/340294/.

Gyawali, B., Pontika, N. et Knoth, P. 2020. Open Access 2007 - 2017: Country and University Level Perspective. In: Proceedings of the ACM/IEEE Joint Conference on Digital Libraries in 2020. JCDL '20. New York, NY, USA: Association for Computing Machinery, p. 381–384. Disponible sur: https://doi.org/10.1145/3383583.3398606 [Consulté le: 23 novembre 2020].

Houghton, J.W. (2001). Crisis and transition: the economics of scholarly communication. *Learned Publishing* **14**(3):167–176. doi: https://doi.org/10.1087/095315101750240412.

Jonchère, L. (2013). Synthèse sur les politiques institutionnelles de libre accès à la recherche.

Koutras, N. 2020. Open Access Publishing in the European Union: The Example of Scientific Works. *Publishing Research Quarterly* 36(3), p. 418-436. doi: 10.1007/s12109-020-09745-x.

Koutras, N. 2016a. The concept of intellectual property: From Plato's views to current copyright protection in the light of open access. Intellectual Property Forum: journal of the Intellectual and Industrial Property Society of Australia and New Zealand (106), p. 43.

Koutras, N. 2016b. The Desirability of Open Access as a Means of Publication and Dissemination of Information: Time to Recast the Relationship between Commercial Publishers and Authors. University of Western Australia Law Review 41, p. 85.

Kozak, M. and Hartley, J. (2013). Publication fees for open access journals: Different disciplines—different methods. *Journal of the American Society for Information Science and Technology* **64**(12):2591–2594. doi: https://doi.org/10.1002/asi.22972.Laakso, M., & Björk, B.-C. (2012). Anatomy of open access publishing: A study of longitudinal development and internal structure. BMC Medicine, 10(124), 1–9. https://doi.org/10.1186/1741-7015-10-124

Lomazzi, L. and Chartron, G. (2014). The implementation of the European Commission recommendation on open access to scientific information: Comparison of national policies. *Information Services & Use* **34**(3–4):233–240. doi: https://doi.org/10.3233/ISU-140743.

Maddi, A. (2020). Measuring open access publications: a novel normalized open access indicator. *Scientometrics*. doi: https://doi.org/10.1007/s11192-020-03470-0.

McVeigh, M. (2004). Open access journals in the ISI citation databases: Analysis of impact factors and citation patterns – A citation study from Thomson Scientific (Rep.). Thomson Scientific. Retrieved from http://science.thomsonreuters.com/m/pdfs/openaccesscitations2.pdfMescheba, W., Miotti, E.L. et Sachwald, F. 2019. Measuring changes in country scientific profiles: the inertia issue. In: ISSI 2019 Conference Proceedings. Sapienza University, Rome, Italy, p. 1519-1530.

Okpala, H.N. (2017). *Access Tools And Services To Open Access: DOAR, ROAR, SHERPA-RoMEO, SPARC AND DOAJ.Informatics Studies*. Available at: http://eprints.rclis.org/32498/ [Accessed: 18 May 2020].

OpenAIRE, Norway (2020). *Norway*. Available at: https://www.openaire.eu/item/norway [Accessed: 18 March 2020].

Pinfield, S., Salter, J., Bath, P.A., Hubbard, B., Millington, P., Anders, J.H.S. and Hussain, A. (2014). Open-access repositories worldwide, 2005–2012: Past growth, current characteristics, and future possibilities. *Journal of the Association for Information Science and Technology* **65**(12):2404–2421. doi: https://doi.org/10.1002/asi.23131.

Piwowar, H., Priem, J., Larivière, V., Alperin, J., Matthias, L., Norlander, B., Haustein, S. (2018). The state of OA: A largescale analysis of the prevalence and impact of open access articles. PeerJ, 6, e4375. https://doi.org/10.7717/peerj.4375

Rabow, I. and Hedlund, T. (2007). Open Access in the Nordic Countries.

Robinson-Garcia, N., Leeuwen, T.N. van et Torres-Salinas, D. 2020. Measuring Open Access Uptake: Data Sources, Expectations, and Misconceptions. Scholarly Assessment Reports 2(1), p. 15. doi: 10.29024/sar.23.

Rousseau, R. 2018. The F-measure for Research Priority. Journal of Data and Information Science 3(1), p. 1-18. doi: 10.2478/jdis-2018-0001.

Sahoo, J., Birtia, T. & Mohanty, B. (2017). Open access journals in library and information science: A study on DOAJ. International Journal of Information Dissemination and Technology, 7(2), 116-119.

Solomon, D.J., Laakso, M. and Björk, B.-C. (2013). A longitudinal comparison of citation rates and growth among open access journals. *Journal of Informetrics* **7**(3):642–650. doi: https://doi.org/10.1016/j.joi.2013.03.008.

Tananbaum, G. (2003). Of wolves and and boys: the scholarly communication crisis. *Learned Publishing* **16**(4):285–289. doi: https://doi.org/10.1087/095315103322422035.

Tate, D. (2019). CILIP Scotland 2019: Open Access, Plan S and New Models for Academic Publishing.

Waltman, L. 2016. A review of the literature on citation impact indicators. Journal of Informetrics 10(2), p. 365-391. doi: 10.1016/j.joi.2016.02.007.

Zhu, Y. (2017). Who support open access publishing? Gender, discipline, seniority and other factors associated with academics' OA practice. *Scientometrics* **111**(2):557–579. doi: https://doi.org/10.1007/s11192-017-2316-z.